\documentclass[useAMS,usenatbib]{mn2e_moro}

\usepackage[dvips]{graphicx}
\usepackage{color}

\title[Redshift evolution of $M_\star-f_{\rm mol}$ relation in $0<z<2$]
{Redshift evolution of stellar mass versus gas fraction relation in $0<z<2$ regime: observational constraint for galaxy formation models}

\author[K. Morokuma-Matsui and J. Baba]
{Kana Morokuma-Matsui$^{1,2}$\thanks{E-mail:kana.matsui@nao.ac.jp; babajn@elsi.jp} and 
Junichi Baba$^{3}$\\ 
$^{1}$Nobeyama Radio Observatory, National Astronomical Observatory of Japan, 462--2 Nobeyama, Minamimaki, Minamisaku, Nagano 384--1305, Japan\\
$^{2}$Chile Observatory, National Astronomical Observatory of Japan, 2--21--1 Osawa, Mitaka, Tokyo 181-8588, Japan\\
$^{3}$Earth-Life Science Institute, Tokyo Institute of Technology, 2--12--1 Ookayama, Meguro-ku, Tokyo, 152--8550, Japan}

\begin{document}

\date{Accepted 1988 December 15. Received 1988 December 14; in original form 1988 October 11}

\pagerange{\pageref{firstpage}--\pageref{lastpage}} \pubyear{2002}

\maketitle

\label{firstpage}

\begin{abstract}
We investigate the redshift evolution of the molecular gas mass fraction ($f_{\rm mol} = \frac{M_{\rm mol}}{M_\star+M_{\rm mol}}$, where $M_{\rm mol}$ is molecular gas mass and $M_\star$ is stellar mass) of galaxies in the redshift range of $0 < z < 2$ as a function of the stellar mass by combining CO literature data.
We observe a stellar-mass dependence of the $f_{\rm mol}$ evolution where massive galaxies have largely depleted their molecular gas at $z=1$, whereas the $f_{\rm mol}$ value of less massive galaxies drastically decreases from $z=1$.
We compare the observed $M_\star-f_{\rm mol}$ relation with theoretical predictions from cosmological hydrodynamic simulations and semi-analytical models for galaxy formation.
Although the theoretical studies approximately reproduce the observed mass dependence of $f_{\rm mol}$ evolution, they tend to underestimate the $f_{\rm mol}$ values, particularly of less massive ($<10^{10}$ M$_\odot$) and massive galaxies ($>10^{11}$ M$_\odot$) when compared with the observational values.
Our result suggests the importance of the feedback models which suppress the star formation while simultaneously preserving the molecular gas in order to reproduce the observed $M_\star-f_{\rm mol}$ relation.

\end{abstract}


\section{Introduction}
\label{sec:Introduction}

Current galaxy-formation models are constructed so as to reproduce the observed {\it stellar} mass function (SMF) of galaxies.
The statistical properties of the stellar components of galaxies have been investigated by means of large surveys in the optical and near-infrared wavelengths \citep[e.g.][]{Cole:2001df,Kochanek:2001xu}.
It is known that the galaxy formation models based on the canonical cold dark matter hypothesis (the so-called $\Lambda$CDM hypothesis) overpredict the number densities of less massive and massive galaxies when compared with the observed SMFs of galaxies \citep[e.g.][]{Benson:2003qe}.
To resolve this discrepancy, several feedback processes have been proposed to blow-out and/or heat up the cold-gas component, which is the raw material for star formation \citep[e.g.][]{Silk:2012lr,Somerville:2014if}.
For low-mass galaxies, the feedback processes of the blowing-out of the cold gas by supernovae \citep[SNe; e.g.][]{Larson:1974la,Dekel:1986oj}, photoionisation by cosmic ultraviolet background radiation \citep[e.g.][]{Ikeuchi:1986rm,Efstathiou:1992eq,Navarro:1997fq,Gnedin:2000vn,Okamoto:2008rz}, and preheating of the intergalactic medium (IGM) by gravitational pancaking \citep[e.g.][]{Mo:2005sf,Lu:2015ty} have been considered, while for high-mass galaxies, active galactic nucleus (AGN) feedback \citep[e.g.][]{Croton:2006rt} and gravitational heating \citep[e.g.][]{Rees:1977bs,Khochfar:2008wo} have been considered.

Galaxy formation is a complex process which involves {\it both} stars and gas, and therefore, galaxy formation models are required to reproduce the observed properties of not only stellar but also the cold-gas components of galaxies.
The observed mass-metallicity relation of galaxies implies that a galaxy is not a closed-box system, but evolves with the continual cycling of baryons between galaxies and the IGM \citep[e.g.][]{Tremonti:2004rf}.
Such a picture of galaxy evolution is supported by recent cosmological hydrodynamic simulations \citep{Dave:2011rm,Dave:2011zl}.
In such a galaxy-IGM ecosystem context, galaxies are described as a slowly evolving system \citep[i.e. `quasi-equilibrium' model;][]{Bouche:2010kq,Dave:2012zp}, where the galaxies acquire gas from cosmic large-scale filamentary structures \citep[e.g.][]{Keres:2005zl,Dekel:2009qf} which activate star formation, lose gas by strong and ubiquitous outflows which regulate star formation, and again acquire the gas returned from the halo \citep[i.e. wind recycling;][]{Oppenheimer:2010wq}.
Therefore, investigations of the evolution of the gas component in galaxies are essential to understand galaxy formation.

Observations of cold-gas components in higher-redshift galaxies are currently underway.
The cold-gas components, atomic hydrogen gas (${\rm H_I}$) and molecular hydrogen gas (H$_2$), are traced by the 21-cm hyperfine structure line and the 3-mm line from the carbon monoxide (CO) rotational transition, respectively.
Although most ${\rm H_I}$ emission studies\footnote{The absorption line observations of neutral hydrogen are also used to test galaxy formation theories \citep{Rahmati:2013eu,Rahmati:2013qf,Rahmati:2014lq,Rahmati:2015bh,Bird:2014fr,Berry:2014zr}.} in high redshift have been limited to the $z\approx0.1-0.2$ regime due to instrumental frequency coverage and sensitivity \citep{Catinella:2008ng,Freudling:2011qm,Fernandez:2013bq,Rhee:2013bx}, emissions from CO have been detected farther out to $z\sim6$; these emissions have been used to measure the H$_2$ mass in the high-$z$ universe \citep[and references therein]{Carilli:2013oq}.
Recent CO surveys of both local and high-$z$ galaxies have revealed molecular gas fractions as a function of stellar mass \citep{Saintonge:2011hl,Tacconi:2013qd}.
In addition to these direct measurements of cold gas, newer advances in the observational studies of cold-gas components are also expected from other indirect methods based on the star-formation-rate--molecular-gas relation \citep[e.g.][]{Popping:2012gd}, dust continuum \citep{Magdis:2012dn,Santini:2014rw,Scoville:2014rc,Scoville:2015qv}, and the spectral features in the optical band \citep{Morokuma-Matsui:2015ij}.
These recent developments, as well as observations made with the newest generation of radio and sub-mm instruments such as ALMA (Atacama Large Millimeter/submillimeter Array), SKA (Square Kilometer Array), ASKAP (Australian Square Kilometre Array Pathfinder), MeerKAT, and NOEMA (NOrthern Extended Millimeter Array) are expected to provide us important constraints on galaxy formation theories.

Recent cosmological galaxy formation models contain sufficient physical ingredients (e.g. gas cooling, H$_2$ formation, star formation, and feedback) such that direct comparison with observations is now possible.
Although the statistical properties of galaxies have been studied with semi-analytical models \citep[SAMs; e.g.][]{Kauffmann:1999dz,Somerville:1999om,Cole:2000wu,Nagashima:2005lr}, recent large-scale cosmological $N$-body/hydrodynamic simulations also allow us to statistically study galaxy formation and evolution \citep[e.g.][]{Schaye:2010zl,Schaye:2015vn,Okamoto:2014rz,Thompson:2014yg,Vogelsberger:2014rm}.
In addition, some of these theoretical studies have implemented the equilibrium or non-equilibrium formation of H$_2$, as well as star formation from such H$_2$ gas, in cosmological simulations \citep{Gnedin:2009lk,Christensen:2012lq,Thompson:2014yg,Tomassetti:2015ft} and SAMs \citep{Lagos:2011dz,Fu:2012hb,Popping:2014zm,Somerville:2015ly}.

However, these theoretical models still require many `sub-grid' parameters for star formation and feedback models due to limited spatial/mass resolutions and a lack of the full understanding of related physical processes \citep[for more details, see a recent review by][]{Somerville:2014if}.
As described above, there are fewer studies on the evolution of the cold-gas component of galaxies than those on stellar components. 
Recent developments in both observational and theoretical studies on the galactic cold-gas components provide us a unique opportunity to compare theoretical and observational results and to obtain the constraints for theoretical models.

In this study, we investigate as to whether the current galaxy formation models reproduce the observed properties of the cold-gas components of galaxies by comparing the observed and theoretically predicted evolutions of the molecular gas mass fraction with respect to the total baryonic mass as a function of stellar mass.
The outline of the paper is as follows.
We first investigate the redshift evolution inferred from the observational data in section \ref{sec:ObservedEvolution}.
In section \ref{seq:ComparisonTheory}, the observed redshift evolution is compared with the model predictions. 
Further, we discuss feedback models implemented in galaxy formation models in section \ref{sec:DiscrepancyPossibleCausesModels}.
Finally, we summarise the study and comment on possible future prospects in this direction in section \ref{sec:Summary}.

\section{Observed Evolution of Molecular Gas Fraction}
\label{sec:ObservedEvolution}

In this section, we investigate the observed molecular gas fraction as a function of stellar mass.
The molecular gas mass fraction is defined as 
\begin{equation}
f_{\rm mol}=\frac{M_{\rm mol}}{M_{\rm mol}+M_\star},
\end{equation}
where $M_\star$ and $M_{\rm mol}$ denote the stellar mass and molecular gas mass of a galaxy, respectively.
We combine the literature data of $M_\star$ and $M_{\rm mol}$ and calculate the molecular gas mass fraction $f_{\rm mol}$ of various galaxies.
The literature data of $M_{\rm mol}$ include both direct and indirect estimations.
In the following sections, we first describe the literature data used in our study (sections \ref{sec:LiteratureCO} and \ref{sec:IndirectMethod}) and we subsequently investigate the redshift evolution of $f_{\rm mol}$ as a function of stellar mass (section \ref{sec:Downsizing}).

\subsection{Measurement of molecular gas mass from CO}
\label{sec:LiteratureCO}

\begin{table*}
\begin{center}
\begin{tabular}{lccccc}
\hline\hline
Name & Redshift $z$& $M_\star$ {\footnotesize ($10^{10}M_\odot$)} & SFR {\footnotesize ($M_\odot/{\rm yr}$)} & Symbols in figure \ref{fig:mstar_fmol_AllFB1} & Reference \\
\hline
Ler08 & $ 0.0007-0.004$ & $0.001-8.0$ & $0.7-4.1$ & open suquare & \citet{Leroy:2008fb}\\
Bos14 & $ 0.0035-0.006$ & $0.03-13.1$ & $0.01-6.0$ & cross & \citet{Boselli:2014yq}\\
Bot14 & $ 0.01-0.03$ & $0.03-1.0$ & $0.16-4.0$ & open pentagon & \citet{Bothwell:2014lr}\\
Sai11 & $0.025-0.05$ & $1.0-32$ & $0.07-40$ & dot & \citet{Saintonge:2011hl}\\
Bau13 & $0.05-0.3$ & $4.0-30$ & $3.4-88$ & filled suquare & \citet{Bauermeister:2013qr}\\
Mor15 & $0.1-0.2$ & $4.0-20$ & $8.5-48$ & filled circle & \cite{Morokuma-Matsui:2015ij}\\
Gea11 & $0.4$ & $4.1-11$ & $28-62$ & open circle & \citet{Geach:2011uo}\\
Dad10 & $1.5$ & $3.3-11$ & $62-400$ & open triangle & \citet{Daddi:2010qq}\\
Tac13 & $1.0-1.5$, $2.0-2.3$ & $0.6-17$ & $26-480$ & filled triangle & \citet{Tacconi:2013qd}\\
Pop12 & $0.5-2.0$ & $0.01-100$ & $-$ & dash-dot lines & \citet{Popping:2012gd}\\
\hline
\end{tabular}
\caption{Summary of samples from the literature. Estimations of $M_\star$ and SFR in each study are described in section \ref{sec:LiteratureCO}.}
\label{tab:PreviousStudies}
\end{center}
\end{table*}

CO data were retrieved from nine studies \citep{Leroy:2008fb,Daddi:2010qq,Geach:2011uo,Saintonge:2011hl,Bauermeister:2013qr,Tacconi:2013qd,Boselli:2014yq,Bothwell:2014lr,Morokuma-Matsui:2015ij} to cover a wide range of redshifts and stellar masses.
We adopted the Galactic conversion factor of $\alpha_{\rm CO}=4.35$ M$_\odot$(K km s$^{-1}$ pc$^2$)$^{-1}$ \citep[including the contribution of helium;][]{Bolatto:2013rr} for all the sample galaxies\footnote{The starburst galaxies, which require different $\alpha_{\rm CO}$, are excluded in this study even though they were included in the original studies.} considered in this study to estimate the molecular gas mass as
\begin{equation}
M_{\rm mol} = \alpha_{\rm CO} L'_{\rm CO},
\end{equation}
where 
$L'_{\rm CO}$ represents the luminosity of the $^{12}$CO($J=1-0$) line in units of K km s$^{-1}$ pc$^2$.
The CO data of \citet{Leroy:2008fb}, \citet{Bothwell:2014lr}, \citet{Tacconi:2013qd}, and \citet{Daddi:2010qq} are not based on $^{12}$CO($J=1-0$) observations but based on $^{12}$CO($J=3-2$) or $^{12}$CO($J=2-1$).
The conversions to $^{12}$CO($J=1-0$) intensity assumed in these studies are described in the following lists.
We discuss effects of variable $\alpha_{\rm CO}$ on $f_{\rm mol}$ in section \ref{sec:ComparisonObservationTheory}.
The literature data of the CO observations are summarised in Table \ref{tab:PreviousStudies}.
We also describe the stellar mass and star formation rate (SFR) estimations of these literature data in the following list.

\begin{itemize}
\item \citet{Leroy:2008fb}:
Leroy et al. combined the literature data of the CO maps of local galaxies ($3<D<30$ Mpc) including spiral and dwarf galaxies, $^{12}$CO($J=2-1$) data from HERACLES \citep{Leroy:2008fb} and $^{12}$CO($J=1-0$) data from BIMA SONG \citep{Helfer:2003kx}.
They assumed a $^{12}$CO($J=2-1$) to $^{12}$CO($J=1-0$) intensity ratio of 0.8.
The stellar mass and SFR of the sample galaxies range over $10^{7.1}-10^{10.9}$ M$_\odot$ and $0.7-4.1$ M$_\odot$ yr$^{-1}$, respectively and they are the so-called main sequence of star-forming galaxies\footnote{
It is known that star-forming galaxies form a distinct sequence of SFR with $M_\star$\citep[e.g.][]{Noeske:2007lr}.
The galaxies on this sequence are called `main sequence' galaxies.}.
Leroy et al. derived stellar mass using Spitzer's 3.6-$\mu$m data and SFR using GALEX FUV and Spitzer's 24-$\mu$m data.

\item \citet{Boselli:2014yq}:
Boselli et al. observed $^{12}$CO($J=1-0$) emission data of 59 late-type galaxies of the Herschel Reference Survey, which is a survey of the complete $K-$band selected, volume-limited ($15<D<25$ Mpc) galaxies spanning a wide range in morphological type and luminosity, using the NRAO Kitt Peak 12-m telescope.
They combined the literature data of CO and provided a CO catalogue of 225 out of the 322 galaxies of their complete sample with stellar mass of $10^{8.5}-10^{11.2}$ M$_\odot$ and SFR of $0.01-6$ M$_\odot$ yr$^{-1}$.
The stellar masses of their samples were estimated in \citet{Cortese:2012oq} from $i-$band luminosities using the $g-i$ colour-dependent stellar mass-to-luminosity ratio relation from \citet{Zibetti:2009eu}.
The sample galaxies are the main sequence of star-forming galaxies \citep{Cortese:2014qe,Ciesla:2014lr}.

\item \citet{Bothwell:2014lr}:
Bothwell et al. presented the first data release of a $^{12}$CO($J=2-1$) survey with the 12-m telescope of Atacama Pathfinder EXperiment (APEX) for nearby dwarf galaxies at $0.01<z<0.03$ with the stellar mass and SFR ranges of $10^{8.5}-10^{10}$ M$_\odot$ and $0.1-4.0$ M$_\odot$ yr$^{-1}$, respectively.
They limited to the samples with metallicity of $12+\log{(\rm O/H)}>8.5$ for $\alpha_{\rm CO}$ to be similar to or lower than the Milky Way value.
They assumed a $^{12}$CO($J=2-1$) to $^{12}$CO($J=1-0$) intensity ratio of unity to estimate $L'_{\rm CO}$.
The stellar masses and SFRs of their sample galaxies were provided by the Max Planck Institute for Astrophysics-John Hopkins University (MPA-JHU) group.
The stellar masses of the galaxies were derived by fitting the SDSS $ugriz$ photometry to models spanning a wide range of star formation histories.

\item \citet{Saintonge:2011hl}: 
Saintonge et al. conducted a CO survey of local galaxies with stellar masses of $(1-30)\times10^{10}$ M$_\odot$ and SFRs of $0.07-40$ M$_\odot$ yr$^{-1}$ with the 30-m telescope at the IRAM facility (CO Legacy Data base for the GASS survey, COLD GASS), and their targets were selected such that the resulting stellar mass distribution was roughly flat.
In their study, the stellar mass and SFR of galaxies were estimated by fitting the GALEX (NUV and FUV) and SDSS photometries with the \citet{Bruzual:2003ak} population synthesis code \citep{Saintonge:2011kx}.
Their sample consists of the star forming galaxies and passive galaxies.

\item \citet{Bauermeister:2013qr}: 
Bauermeister et al. targeted 31 star-forming galaxies with stellar masses in the range of $(4-30)\times10^{10}$ M$_\odot$ and SFRs in the range of $3-90$ M$_\odot$ yr$^{-1}$ at $z\sim0.05-0.5$ and detected CO emissions from 24 galaxies using the Combined Array for Research in Millimeter-wave Astronomy (CARMA) facility (Evolution of molecular Gas in Normal Galaxies, EGNoG).
The sample galaxies in the ranges of $z\sim0.05-0.32$ from which CO emission was detected were drawn from SDSS DR7 \citep{Abazajian:2009cz}.
The stellar mass and SFR of their sample galaxies were provided by the MPA-JHU group.
This estimate was found to be comparable with the one obtained from stellar absorption features \citep{Kauffmann:2003vh}.
Further, SFRs were derived by fitting the fluxes of no less than five emission lines \citep{Brinchmann:2004sy}.
In this study, we only use the data of the normal star-forming galaxies of their sample.

\item \citet{Morokuma-Matsui:2015ij}:
Morokuma-Matsui et al. conducted CO observations of 12 galaxies with stellar masses of $(4-20)\times10^{11}$ M$_\odot$ and SFRs of $8.5-48$ M$_\odot$ yr$^{-1}$ in the $z\sim0.1-0.2$ regime using the 45-m telescope at the Nobeyama Radio Observatory, and they detected CO emission from 8 galaxies.
Similar to the case of \citet{Bothwell:2014lr} and \citet{Bauermeister:2013qr}, the stellar masses and SFRs of their samples were also obtained from the MPA-JHU group and drawn from SDSS DR10 \citep{Ahn:2014ys}.
They showed that their samples correspond to star-forming galaxies at the redshift.

\item \citet{Geach:2011uo}:
Geach et al. observed seven galaxies selected based on the basis of the 24-$\mu$m emission galaxies with masses of $(4-11)\times10^{11}$ M$_\odot$ and SFRs of $28-62$ M$_\odot$ yr$^{-1}$ at $z\sim0.4$ using the IRAM Plateau de Bure Interferometer (PdBI) and detected CO emission from five galaxies.
The stellar masses of their samples were estimated by fitting $BVRIJK$ photometry \citep{Moran:2007vn} 
to model spectral energy distributions (SEDs) using the {\tt KCORRECT} software package, v4.2 \citep{Blanton:2007rt}.
The SFRs of their samples were derived from the far-infrared luminosity \citep{Kennicutt:1998hb} estimated 
from the 7.7-$\mu$m-line luminosity \citep{Geach:2009yq}.

\item \citet{Daddi:2010qq}:
Their sample galaxies comprised the star-forming galaxies with a stellar mass range of $(3-11)\times10^{10}$ M$_\odot$ and SFR range of $62-400$ M$_\odot$ yr$^{-1}$ in the $1.4<z<2.5$ regime classified by using the $BzK$ colour criterion \citep{Daddi:2004fr} with Spitzer imaging detection at 24 $\mu$m.
They observed CO($J=2-1$) of five galaxies with IRAM PdBI and estimated $L'_{\rm CO}$ by assuming a CO($J=2-1$)-to-CO($J=1-0$) conversion factor of 0.86 \citep{Dannerbauer:2009th}.
The stellar mass was estimated via an empirical method with $BzK$ photometry, which was calibrated 
with the mass estimates from multicolor photometry and SED fitting \citep{Daddi:2004fr}.
The SFRs of their samples were the averages of three SFRs estimated from dust-corrected UV, 24-$\mu$m emission, and 1.4-GHz radio fluxes \citep{Daddi:2010qq}.

\item \citet{Tacconi:2013qd}:
Tacconi et al. conducted the largest CO survey of $z\sim1-2$ galaxies with stellar mass of $(0.6-17)\times10^{10}$ M$_\odot$ and SFR of $26-480$ M$_\odot$ yr$^{-1}$ with the IRAM PdBI and detected CO emissions from 52 galaxies (IRAM Plateau de Bure HIgh-z Blue Sequence Survey, PHIBSS).
The stellar mass was estimated from SED fitting.
The SFR estimate was based on the sum of the observed UV- and IR-luminosities for the $z\sim1.2$ samples \citep{Wuyts:2011uq}, and the extinction-corrected H$\alpha$ luminosities for the $z\sim2.2$ samples 
\citep{Kennicutt:1998hb,Forster-Schreiber:2009fj,Mancini:2011kx,Wuyts:2011uq}.
Their sample follows the main sequence of star-forming galaxies at the redshift.
They observed the luminosities of the $^{12}$CO($J=3-2$) lines of galaxies and converted these data sets into the luminosity of $^{12}$CO($J=1-0$) lines using a $^{12}$CO($J=1-0$)$/^{12}$CO($J=3-2$) ratio of 2.

\end{itemize}
All the studies except for \citet{Leroy:2008fb} assumed the Chabrier initial mass function \citep[IMF,][]{Chabrier:2003oe} for estimating the stellar mass and SFR of galaxies.
\citet{Leroy:2008fb} used Kroupa IMF \citep{Kroupa:2001fj}.
We did not correct for this IMF difference, since the difference in the stellar mass and SFR estimations based on Chabrier and Kroupa IMFs is small compared to the other uncertainties \citep{Chomiuk:2011lr,Kennicutt:2012yq,Madau:2014fk}.

\subsection{Estimation of molecular gas mass from SFR}
\label{sec:IndirectMethod}

We also use the molecular gas mass estimated indirectly from optical studies,
since the CO measurement is limited to galaxies with 
a relatively narrow range of stellar mass ($\sim10^{10}-10^{11}$ M$_\odot$), particularly at high redshift.

\citet{Popping:2012gd} estimated the $f_{\rm mol}$ values of the COSMOS \citep[Cosmological Evolution Survey;][]{Scoville:2007qy} galaxies using the molecular gas mass values inferred from the Kennicutt-Schmidt (K-S) law and the pressure-based H$_2$ formation model \citep{Blitz:2004si,Blitz:2006le}.
They presented the fitting formulae of the $M_\star-f_{\rm mol}$ relations for star-forming galaxies and all galaxies brighter than $I_{\rm AB}=24$ mag.
For the star-forming galaxies, the fitting formula is given by 
\begin{equation}
f_{\rm mol,P12, SF}=\frac{1}{\exp{(\log{M_\star}-A)/B}+1},
\label{eq:fmol_Pop12_SF-1}
\end{equation}
where
\begin{eqnarray}
A=6.15\left(1+\frac{z}{0.036}\right)^{0.144},~~
B=1.47(1+z)^{-2.23},
\label{eq:fmol_Pop12_SF-2}
\end{eqnarray}
and that for the whole galaxies brighter than 24 mag in $I_{\rm AB}$ is
\begin{equation}
f_{\rm mol,P12}=\frac{1}{\exp{(\log{M_\star}-C)/D}+1},
\label{eq:fmol_Pop12-1}
\end{equation}
where 
\begin{eqnarray}
C=6.33\left(1+\frac{z}{0.016}\right)^{0.115},~~
D=1.71(1+z)^{-1.57}.
\label{eq:fmol_Pop12-2}
\end{eqnarray}

Here, we mention that \citet{Popping:2012gd} showed that their $I_{\rm AB}<24$ mag sample is complete ($100$ \%) for the range of $10^{9.2}\ {\rm M_\odot}<M_\star<10^{11.7}\ {\rm M_\odot}$ at $0.5<z<0.75$ with respect to the SMF derived in the same field \citep{Ilbert:2010nr}.
However, it is almost evenly $<50 \%$ for all stellar mass ranges at $z>1.25$.
In addition, if we assume the fitting functions of \citet{Popping:2012gd} as accurate, the $f_{\rm mol}$ values of galaxies with stellar masses $>10^{11}$ M$_\odot$ becomes larger at lower redshifts, but the statistics with regards to higher-$z$ massive galaxies in the fitting is poor.
Thus, it is required to closely consider these observational limitations when we compare observations with theoretical studies.

\begin{figure*}
\includegraphics[width=180mm]{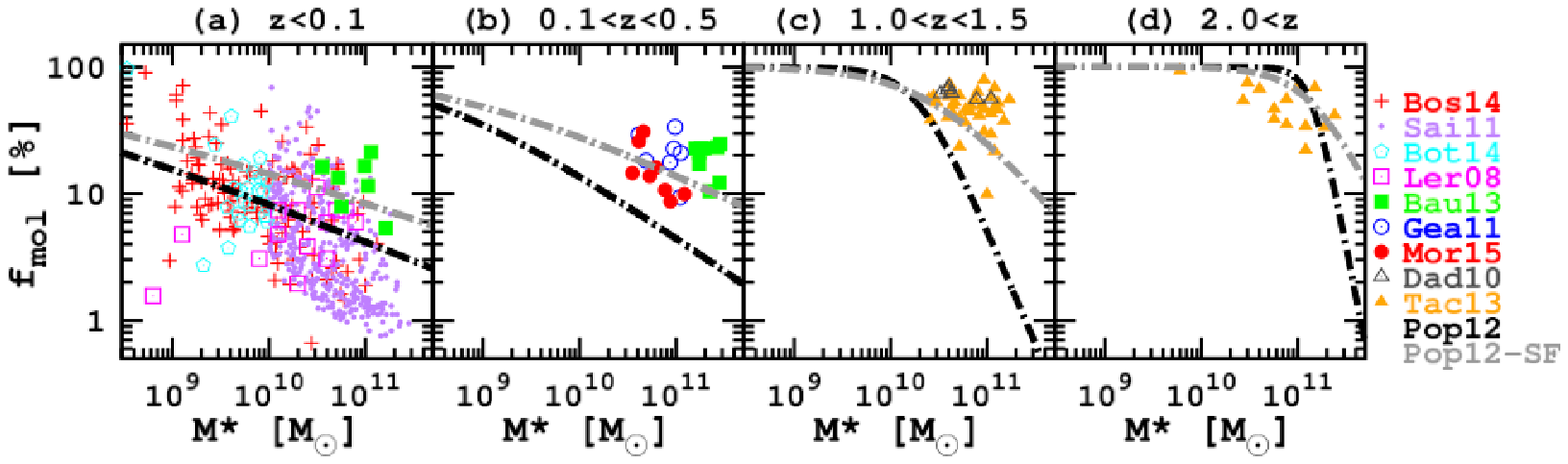}
 \caption{Molecular gas mass fraction $f_{\rm mol}$ as a function of stellar mass and redshift (a: $z<0.1$, b: $0.1<z<0.5$, c: $1.0<z<1.5$ and d: $2.0<z$).
The symbols indicate CO data from the literature (cross: \citet{Boselli:2014yq}, open-square: \citet{Leroy:2008fb}, small-circle: \citet{Saintonge:2011hl}, open pentagon: \citet{Bothwell:2014lr}, filled-square: \citet{Bauermeister:2013qr}, filled-circle: \citet{Morokuma-Matsui:2015ij}, open-circle: \citet{Geach:2011uo}, open-triangle: \citet{Daddi:2010qq}, and filled-triangle: \citet{Tacconi:2013qd}).
The grey and black dot-dash lines indicate $M_\star-f_{\rm mol}$ relations based on the K-S law and pressure-based H$_2$ formation models for the star-forming galaxies ($f_{\rm mol, P12, SF}$) and whole samples ($f_{\rm mol, P12}$), respectively in \citet{Popping:2012gd}.
The redshifts of the equations for $f_{\rm mol, P12, SF}$ and $f_{\rm mol, P12}$ are 0.025 for (a), 0.3 for (b), 1.25 for (c), and 2.2 for (d).
The relationship between the symbols and references is summarised in Table \ref{tab:PreviousStudies}.
The stellar-mass-dependent $f_{\rm mol}$ evolution, where massive galaxies tend to reduce $f_{\rm mol}$ in earlier epochs compared to their less massive counterparts, can be clearly observed (see section \ref{sec:Downsizing} for more details).}
  \label{fig:mstar_fmol_AllFB1}
\end{figure*}

\subsection{Stellar-mass-dependent evolution of molecular gas mass fraction of galaxies}
\label{sec:Downsizing}

In this section, we investigate the stellar mass dependence of $f_{\rm mol}$ evolution using the observational data.
Figure \ref{fig:mstar_fmol_AllFB1} shows the redshift evolution of the fraction $f_{\rm mol}$ as a function of stellar mass.
Although the stellar mass range of galaxies with CO measurements is narrow at high redshift ($z>0.1$), we can observe the stellar-mass dependence where the massive galaxies have smaller $f_{\rm mol}$ values than their less massive counterparts at $z\sim0-2$.
In all redshift ranges, the observed $M_\star-f_{\rm mol,CO}$ relations are well reproduced by the $M_\star-f_{\rm mol,P12, SF}$ relations in which the molecular gas mass is estimated from the indirect method.
\citet{Popping:2012gd} have already reported that the M$_\star$--$f_{\rm mol, P12, SF}$ relation succeeds in 
reproducing the observed M$_\star$--$f_{\rm mol, CO}$ relations of local galaxies \citep{Saintonge:2011hl} 
and $z\sim1-2$ samples \citep{Tacconi:2010fe}.
Here, we found that their M$_\star$--$f_{\rm mol, P12, SF}$ relation also reproduces the observed M$_\star$--$f_{\rm mol, CO}$ relations of galaxies at $z\sim 0.1-1$.

There is a clear stellar-mass dependence in the $f_{\rm mol}$ evolution of star-forming galaxies where less massive galaxies tend to show a decrease in $f_{\rm mol}$ over time as reported in \citet{Popping:2012gd,Popping:2015fu}.
Here, we remark that recent dust-based studies have also suggested that the stellar-mass dependence of the cold gas mass fraction $(f_{\rm gas} =\frac{M_{\rm gas}}{M_{\rm gas}+M_\star}$, where $M_{\rm gas}$ denotes the total cold gas mass including the atomic and molecular gas) shows a similar stellar-mass dependent evolution \citep{Santini:2014rw}.
Therefore, studies based on two independent methods (CO or dust) both suggest the stellar-mass dependent evolution of the cold gas fraction of galaxies.

We examine the stellar-mass dependent evolution of $f_{\rm mol}$ quantitatively.
We must consider the stellar mass growth when discussing the $f_{\rm mol}$ evolution.
If we adopt the stellar mass growth of star-forming galaxies as reported in \citet{Leitner:2012ph}, 
the redshift evolution of the $M_\star-f_{\rm mol, P12, SF}$ relation suggests that galaxies with stellar masses of $10^{10}$ M$_\odot$, $10^{10.5}$ M$_\odot$, and $10^{11}$ M$_\odot$ at $z=0$ 
exhibit decreases by factors of 8 (from 80 \% to 10 \%), 6 (37 \% to 7.0 \%), and 4 (26 \% to 6.1 \%) in their molecular gas mass fraction from $z=1$ to $z=0$, respectively\footnote{
Here we use the $M_\star-f_{\rm mol, P12, SF}$ relation 
because \citet{Leitner:2012ph} estimated the growth rate of stellar mass by assuming 
that present-day star-forming galaxies were on the `main sequence' of the star-forming galaxies.}.
Similarly, if we use the stellar mass growth of galaxies based on the abundance matching method \citep[i.e. not biased with respect to star-forming galaxies;][]{Conroy:2009cs}, the $M_\star-f_{\rm mol, P12}$ evolution suggests that the galaxies with stellar masses of $10^{10}$, $10^{10.5}$, and $10^{11}$ M$_\odot$ at $z=0$ exhibit molecular gas-fraction decreases by factors of 14 (from 96 \% to 6.8 \%), 8 (14 \% to 4.2 \%), and 1 (4.9 \% to 3.6 \%) from $z=1$ to $z=0$, respectively.
Therefore, the $f_{\rm mol}$ evolution of less massive galaxies in the range of $z=0-2$ is more drastic than 
that of massive galaxies, and the massive galaxies appear to have depleted relatively large fractions of their molecular gas by $z=1$\footnote{
$f_{\rm mol}$ evolution farther out to $z=3$ has been also investigated \citep{Magdis:2012jt,Saintonge:2013pi}.
\citet{Magdis:2012jt} detected a CO($J=3-2$) emission from an infrared luminous Lyman break galaxy at $z=3.2$ 
with a stellar mass of $2\times10^{11}M_\odot$.
The estimated $f_{\rm mol}$ is 36 \% suggesting that massive galaxies have flat evolution of $f_{\rm mol}$ at least in $1<z<3$.
}
\citep{Popping:2015fu}.

\section{Comparison with theoretical studies}
\label{seq:ComparisonTheory}

In this section, we compare the observed $M_\star-f_{\rm mol}$ relation with predictions of theoretical studies constructed to reproduce the observed SMF.
The theoretical studies we used in this study are briefly summarised in section \ref{sec:LiteratureModels}.
We demonstrate a discrepancy in the $M_\star-f_{\rm mol}$ relation between the observations and theoretical predictions in section \ref{sec:ComparisonObservationTheory}.

\begin{figure*}
\includegraphics[width=180mm]{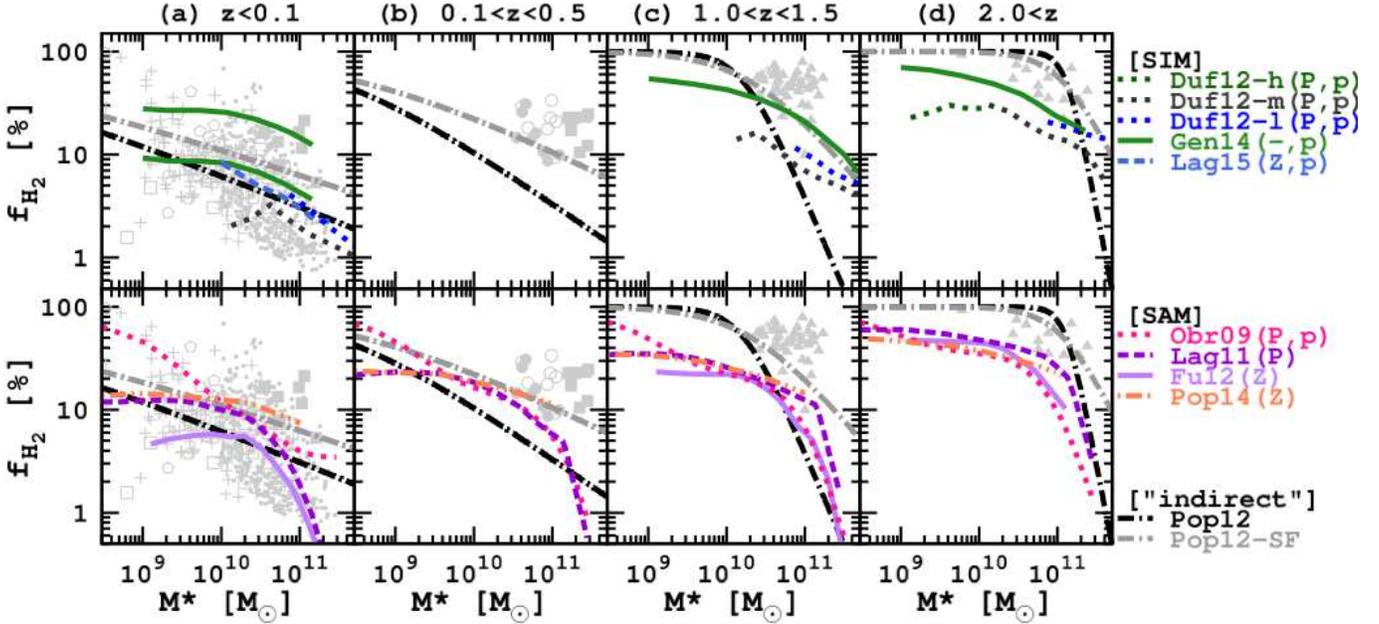}
 \caption{H$_2$ gas mass fraction $f_{\rm H_2}$ as a function of stellar mass and redshift (a: $z<0.1$, b: $0.1<z<0.5$, c: $1.0<z<1.5$ and d: $2.0<z$).
 The black and grey dot-dash lines indicate the $M_\star-f_{\rm H_2}$ relations based on the K-S law and pressure-based H$_2$ formation models for the star forming galaxies ($f_{\rm H_2, P12, SF}$) and whole samples ($f_{\rm H_2, P12}$), respectively in \citet{Popping:2012gd}.
The redshifts of $f_{\rm H_2, P12, SF}$ and $f_{\rm H_2, P12}$ are 0.025 for (a), 0.3 for (b), 1.25 for (c), and 2.2 for (d).
[Upper panel] Comparison with the predictions from cosmological $N$-body/hydrodynamic simulations (SIMs; dark-green, dark-grey, and blue dotted lines: high-, middle-, and low- resolution calculations of \citet{Duffy:2012vl}, respectively, solid green lines: \citet{Genel:2014fk}, and light-blue dashed lines: \citet{Lagos:2015fj}).
\citet{Genel:2014fk} provided both lower and upper limits for $f_{\rm H_2}$ for galaxies at $z\sim0$.
[Lower panel] Comparison with the predictions from semi-analytic models (SAMs; pink dotted lines: \citet{Obreschkow:2009hi}, purple dashed lines: \citet{Lagos:2011qy}, light-purple solid line: \citet{Fu:2012hb}, orange dot-dot-dashed line: \citet{Popping:2014zm}).
In the graph legend, `P' denotes the pressure-based H$_2$ model, `Z' denotes the metallicity-based H$_2$ model, and `p' indicates that the H$_2$ model was implemented as a post-process.
 The relationship between the each symbol and the reference is summarized in Table \ref{tab:TheoreticalStudies}.
On each plot, we indicate the theoretical predictions for $f_{\rm H_2}$ at $z=0$ in (a), $z=0.5$ in (b), $z=1$ in (c), and $z=2$ in (d).
Theoretical studies `qualitatively' reproduce the stellar mass dependence of $f_{\rm H_2}$ evolution but there is a `non-negligible' discrepancy between the observations and theoretical studies (see section \ref{sec:ComparisonObservationTheory} for details).}
  \label{fig:mstar_fmol_AllFB2}
\end{figure*}

\begin{table*}
\begin{center}
\begin{tabular}{lcccccc}
\hline\hline
Label & Method & H$_2$ prescription & Star formation model & Feedbacks & Line Styles in Fig. \ref{fig:mstar_fmol_AllFB2} & Reference \\
\hline
Duf12 & SIM & Pressure$^\ast$ & \citet{Schaye:2008yf} & SN & dot &\citet{Duffy:2012vl}\\
Gen14 & SIM & --$^\ast$ & \citet{Springel:2003ai} & SN \& AGN & solid & \citet{Genel:2014fk}\\
Lag15 & SIM & Metallicity$^\dagger$$^\ast$ & \citet{Schaye:2008yf} & SN \& AGN & dash & \citet{Lagos:2015fj}\\
Obr09 & SAM & Pressure$^\ast$ & \citet{Kennicutt:1989pi} & SN \& AGN & dot & \citet{Obreschkow:2009hi}\\
Lag11 & SAM & Pressure & \citet{Bigiel:2008rt} & SN \& AGN & dash & \citet{Lagos:2011qy,Lagos:2011dz}\\
Fu12 & SAM & Metallicity & \citet{Bigiel:2008rt} & SN \& AGN & solid & \citet{Fu:2012hb} \\
Pop14 & SAM & Metallicity & \citet{Bigiel:2008rt} & SN \& AGN & dot-dot-dash & \citet{Popping:2014zm}\\
\hline
\end{tabular}
\caption{Summary of the theoretical studies used in Fig.~\ref{fig:mstar_fmol_AllFB2}.
Details of each study are described in section \ref{sec:LiteratureModels}.
$^\dagger$The H$_2$ prescription is from \citet{Gnedin:2011wj}, which provided a fitting formula to their galaxy simulations incorporating non-equilibrium chemical network of hydrogen and helium and non-equilibrium cooling and heating rates.
The other `Metallicity' models use an analytical formula based on the chemical equilibrium model of H$_2$ \citep{Krumholz:2009sf}.
$^\ast$H$_2$ formation (partitioning) models were implemented as a post-process.
}
\label{tab:TheoreticalStudies}
\end{center}
\end{table*}

\subsection{Galaxy formation models in the literatures}
\label{sec:LiteratureModels}

Recent galaxy formation models take into account star formation from molecular gas, 
as suggested by observations \citep[e.g.][]{Bigiel:2008rt,Leroy:2008fb}. 
These theoretical models need to consider both star formation and molecule formation.  
Since star formation is an unknown process and the current cosmological numerical simulations of galaxy formation cannot resolve the spatial scale of molecular clouds, 
the galaxy formation models treat star formation as a `sub-grid' model (see below).

For H$_2$ formation, most hydrodynamic simulations and SAMs adopt simplified methods such as `pressure-based' and `metallicity-based' models, although some cosmological hydrodynamic simulations solve the non-equilibrium chemical reactions of hydrogen \citep{Gnedin:2009lk}.
The pressure-based model is an empirical method based on the observed relation 
between the molecule fraction of cold gas and the surface densities of stellar and gas components 
in nearby disc galaxies \citep{Blitz:2004si,Blitz:2006le,Leroy:2008fb}.
In contrast, the metallicity-based model uses an analytical formulae of the molecule fraction as a function of the gas density, metallicity, and radiation field \citep{Krumholz:2008ez,Krumholz:2009sf}.
It has been shown that both these types of models suitably reproduce the observed molecular fraction of cold gas well \citep{Blitz:2004si,Blitz:2006le,Krumholz:2008ez,Fumagalli:2010uq}.

In this study, we compare the observations to the predicted $M_\star-f_{\rm mol}$ relations 
from three cosmological $N$-body/hydrodynamic simulations \citep[SIM;][]{Duffy:2012vl,Genel:2014fk,Lagos:2015fj} and four SAMs \citep{Obreschkow:2009hi,Lagos:2011qy,Fu:2012hb,Popping:2014zm} of galaxy formation.
We summarise the theoretical studies used in this study in Table \ref{tab:TheoreticalStudies}.

\begin{itemize}
\item \citet[SIM]{Duffy:2012vl}:
Duffy et al. implemented the pressure-based H$_2$ formation model \citep{Blitz:2006le,Leroy:2008fb}
in the calculations of the OverWhelmingly Large Simulations (OWLS) project \citep{Schaye:2010zl} as a post-process.
They determined the free parameters of the H$_2$ formation model according to recent observations \citep{Leroy:2008fb}.
Their estimated H$_2$ mass included the contribution of helium.
The simulations in the OWLS include radiative cooling \citep[$10^4~{\rm K}< T < 10^8~{\rm K}$][]{Wiersma:2009lr}, photoionisation heating from cosmic UV background radiation, star formation \citep{Schaye:2008yf}, mechanical feedback from SNe \citep[i.e. by `kick';][]{Dalla-Vecchia:2008fk} and thermal feedback from AGN \citep{Booth:2009la}.
Star formation is based on the study by \cite{Schaye:2008yf}, in which they analytically recast the K-S law as a function of pressure rather than density, while assuming a self-gravitating disc.
Duffy et al. provided results for the different co-moving box sizes: 25 $h^{-1}$ Mpc (high resolution), 50 $h^{-1}$ Mpc (middle resolution), and 100 $h^{-1}$ Mpc (low resolution).
The simulations with the smallest box size were run to $z=2$ and those with the larger box sizes were run to $z=0$.
Although \citet{Duffy:2012vl} examined both cases of with and without AGN feedback,
we adopted the former case since Duffy et al. already reported that the latter results in a large discrepancy between the simulation and observation.

\item \citet[SIM]{Genel:2014fk}
Genel et al. did not implement any H$_2$ formation models but assumed that one-third of the neutral gas of the galaxy is in the molecular phase to derive the lower limit for $f_{\rm H_2}$ of $z\sim0$ galaxies formed in the Illustris simulation \citep{Vogelsberger:2013kl}.
Their estimated H$_2$ mass does not include the contribution of helium.
They also assumed that all the neutral gas in a galactic disc is in the molecular phase to derive $f_{\rm H_2}$ of high-$z$ galaxies and the upper limit of $z\sim0$ galaxies.
The Illustris uses a moving mesh code, AREPO \citep{Springel:2010bs} and includes radiative cooling, heating from time-dependent UV background radiation, star formation \citep{Springel:2003ai}, mechanical feedback from SNe and  thermal/mechanical/radiative feedbacks from AGNs \citep{Springel:2005dp,Sijacki:2007tg,Vogelsberger:2013kl}.

\item \citet[SIM]{Lagos:2015fj}
Lagos et al. implemented H$_2$ formation models as a post-process in EAGLE \citep[Evolution and Assembly of GaLaxies and their Environments;][]{Schaye:2015vn}, which is a cosmological numerical simulation that uses an $N$-body/SPH code, GADGET-3 \citep{Springel:2008hc}.
Here, H$_2$ formation is modelled according to two prescriptions depending on the local dust-to-gas ratio and the interstellar radiation field \citep{Gnedin:2011wj,Krumholz:2013sp}.
Their estimated H$_2$ mass does not include the contribution of helium.
They considered radiative cooling \citep{Wiersma:2009lr,Schaye:2015vn}, photo-heating by cosmic microwave background, UV and X-ray background radiation \citep{Haardt:2001fy}, star formation \citep{Schaye:2008yf}, and thermal feedbacks from SNe \citep{Dalla-Vecchia:2012fv} and AGN \citep{Booth:2009la}. 
We employed their $M_\star-f_{\rm mol}$ relation \citep[`Ref-L100N1504' model, see Fig. 5 of][]{Lagos:2015fj} calculated using the H$_2$ formation model of \citet{Gnedin:2011wj}.

\item \citet[SAM]{Obreschkow:2009hi}:
They implemented the pressure-based H$_2$ formation (partitioning) model \citep{Blitz:2006le,Leroy:2008fb} as a post-process in the semi-analytical model, L-GALAXIES \citep{Croton:2006rt,De-Lucia:2007ys},
in which the dark matter halo and sub-halo merger trees are constructed from the Millennium Simulation \citep{Springel:2005vn}.
With the L-GALAXIES, they implemented radiative cooling with the cooling function of \citet{Sutherland:1993uq}, photoionisation heating \citep{Gnedin:2000vn,Kravtsov:2004rt}, star formation \citep{Kauffmann:1996uq,Kennicutt:1989pi}, and the feedbacks from SNe and AGN \citep{Croton:2006rt}.
We obtain their $M_{\rm H_2}/M_\star$ values from Fig. 9 of \citet{Popping:2015fu} and recalculate $f_{\rm H_2}$.
Here, we mention that \citet{Popping:2015fu} did not include the contribution from helium.

\item \citet[SAM]{Lagos:2011dz,Lagos:2011qy}:
Lagos et al. implemented both pressure-based \citep{Blitz:2006le} and metallicity-based \citep{Krumholz:2009sf} H$_2$ formation models in the semi-analytical model, GALFORM \citep{Cole:2000wu,Benson:2003qe,Baugh:2005rw,Bower:2006qa}, in which the merger trees of dark matter halo were constructed using the Monte Carlo method \citep{Lacey:1993kl}.
They considered the shock heating and radiative cooling \citep{Sutherland:1993uq} inside the dark matter halos, star formation \citep{Lagos:2011dz,Kennicutt:1998fk,Leroy:2008fb,Bigiel:2008rt,Krumholz:2009sf}, and the feedbacks from SNe and AGN \citep{Baugh:2005rw,Bower:2006qa}.
We obtain their $M_{\rm H_2}/M_\star$ values \citep[pressure-based H$_2$ model and star formation model based on][]{Bigiel:2008rt} from Fig. 9 of \citet{Popping:2015fu} and recalculate $f_{\rm H_2}$.
Here, we mention that \citet{Popping:2015fu} did not include the contribution from helium.

\item \citet[SAM]{Fu:2010yq, Fu:2012hb}:
Fu et al. implemented both pressure-based \citep{Blitz:2006le} and metallicity-based \citep{Krumholz:2009sf} H$_2$ formation models in the L-GALAXIES \citep{Croton:2006rt,De-Lucia:2007ys} as is the case with \citet{Obreschkow:2009hi}.
Unlike in \citet{Obreschkow:2009hi}, H$_2$ formation in this study is calculated in a self-consistent manner.
The mass contribution from helium was taken into account \citep[a factor of 1.4,][]{Arnett:1996dk,Fu:2010yq}.
Four kinds of star formation models were implemented: 1) `Bigiel' \citep{Bigiel:2008rt,Leroy:2008fb}, 2) `Kennicutt' \citep{Kennicutt:1989pi,Kennicutt:1998fk}, 3) `Genzel' \citep{Genzel:2010fj} and 4) `Krumholz' \citep{Krumholz:2009kx} models.
They concluded that the `Bigiel' model results in better agreement with the observed mass--metallicity relation among the four models.
We employ their $M_\star-f_{\rm mol}$ relation calculated using metallicity-based H$_2$ formation and the `Bigiel' star formation model for a comparison purpose in this study.

\item \citet[SAM]{Popping:2014zm}:
Popping et al. implemented pressure-based \citep{Blitz:2006le} and the metallicity-based \citep{Gnedin:2009lk} H$_2$ formation models in the Santa Cruz semi-analytical model \citep{Somerville:1999om,Somerville:2008oz,Somerville:2012mi}, in which the dark matter halo merger tree was constructed using the Extended Press--Schechter formalism \citep{Somerville:1999xr,Somerville:2008lr}.
They considered photoionization heating \citep{Gnedin:2000vn,Kravtsov:2004rt}, radiative cooling \citep{Sutherland:1993uq,Somerville:1999om}, star formation from molecular gas \citep{Bigiel:2008rt}, and feedbacks from SNe and AGN.
We obtain their $M_{\rm H_2}/M_\star$ values \citep[metallicity-based H$_2$ model and star formation model based on][]{Bigiel:2008rt} from Fig. 9 of \citet{Popping:2015fu} and recalculate $f_{\rm H_2}$.
Here, we mention that \citet{Popping:2015fu} did not include the contribution from helium.

\end{itemize}
We use the H$_2$ gas mass fraction $f_{\rm H_2}=\frac{M_{\rm H_2}}{M_{\rm H_2}+M_\star}$, which does not include the contribution form helium, instead of $f_{\rm mol}$ for a comparison between the observations and the theoretical predictions, since most theoretical studies considered here derived $f_{\rm H_2}$.
For the studies presenting $f_{\rm mol}$ instead of $f_{\rm H_2}$, we recalculate $M_{\rm H_2}$ from $M_{\rm mol}$ using a conversion factor of 1.36 for \citet{Duffy:2012vl} and \citet[indirect method]{Popping:2012gd}, and 1.4 for \citet{Fu:2012hb}.
Hereafter, the $f_{\rm H_2}$ calculated based on equation (\ref{eq:fmol_Pop12_SF-1}) and (\ref{eq:fmol_Pop12_SF-2}) is described as $f_{\rm H_2, P12, SF}$ and the one calculated based on equation (\ref{eq:fmol_Pop12-1}) and (\ref{eq:fmol_Pop12-2}) is described as $f_{\rm H_2, P12}$.

\subsection{Discrepancy between observations and theoretical studies}
\label{sec:ComparisonObservationTheory}

Figure \ref{fig:mstar_fmol_AllFB2} compares the observed and predicted $M_\star-f_{\rm H_2}$ relations from theoretical studies.
Here, we adopt $f_{\rm H_2, P12}$ (black dot-dash lines in Fig. \ref{fig:mstar_fmol_AllFB2}) 
as an observational reference for comparison with theoretical studies since the theoretical studies are not limited to star-forming galaxies.
It should be noted that the theoretical studies did not apply any magnitude limits, which were applied to the observations.
This uncertainty is discussed below.
The predicted $M_\star-f_{\rm H_2}$ relations from SIMs \citep{Duffy:2012vl,Genel:2014fk,Lagos:2015fj}
and SAMs \citep{Obreschkow:2009hi,Lagos:2011dz,Lagos:2011qy,Fu:2010yq, Fu:2012hb,Popping:2014zm} are overlaid on the observed relation shown in Fig. \ref{fig:mstar_fmol_AllFB1}.

The theoretical studies qualitatively reproduce the trends of the stellar-mass-dependent evolution of $f_{\rm H_2}$ where the less massive galaxies have higher $f_{\rm H_2}$ values than their massive counterparts; however, there is a `non-negligible' gap between $f_{\rm H_2}$ from observations and theoretical studies \citep{Popping:2015fu}.
The $f_{\rm H_2}$ predictions from the theoretical studies appear to be consistent with the $f_{\rm H_2, P12}$ values in the stellar mass range of $\sim10^{10}-10^{11}$ M$_\odot$ in Fig. \ref{fig:mstar_fmol_AllFB2}a, which is consistent with the results of a recent study \citep{Lagos:2015fj}.
However, the predicted $f_{\rm H_2}$ values by the SAMs of less massive ($<10^{10}$ M$_\odot$) are underestimated with respect to the observed values except for \citet{Obreschkow:2009hi}.
The discrepancy between the observations and theoretical predictions is still observed even if we only consider nearby galaxies (Figs. \ref{fig:mstar_fmol_AllFB2}a, b) in the stellar mass range of $10^{9.2} {\rm M}_\odot <M_\star<10^{11}$ M$_\odot$, wherein the observational completeness is 100 \% (see section \ref{sec:IndirectMethod}).

\citet{Obreschkow:2009hi} predicted higher $f_{\rm H_2}$ values than the observation values for less massive galaxies, even though they used the same code (L-GALAXIES) as \citet{Fu:2012hb}.
However, \citet{Obreschkow:2009hi} implemented H$_2$ formation as a post-process whereas \citet{Fu:2012hb} calculated H$_2$ formation in a self-consistent manner.
This difference may explain the discrepancy between the two models, but \citet{Obreschkow:2009hi} claimed that their galaxies with a stellar mass of $<4\times10^9$ M$_\odot$ were typically found in a halo with less than 100 particles, whose merging history could be followed over only a few time steps \citep{De-Lucia:2007ys}.

There is a difference in the mass dependence of $f_{\rm H_2}$, even though the SIMs and SAMs predict similar redshift evolution of the SMF of galaxies \citep[see Fig. 4 of][]{Somerville:2014if}.
At all the redshift ranges ($0<z<2$), the SIMs predict more flat $M_\star-f_{\rm H_2}$ relations than those of the observations (upper panels of Fig. \ref{fig:mstar_fmol_AllFB2}), whereas the SAMs predict kinks at the characteristic masses of $10^{10}-10^{11}$ M$_\odot$ (lower panels of Fig. \ref{fig:mstar_fmol_AllFB2}).
The difference in the prescriptions of H$_2$ formation is not likely the reason for this difference between the SAMs and SIMs since there is no clear difference among the SAMs depending on whether or not H$_2$ formation is implemented as a post-process (lower panels of Fig. \ref{fig:mstar_fmol_AllFB2}) or depending on the models \citep[pressure- or metallicity-based;][]{Fu:2012hb}. 
Thus, the difference in modelling star formation and feedback may cause this difference.
We discuss this gap between the SIMs and SAMs in section \ref{sec:ComparisonEM}.

In the following paragraphs, we discuss possible causes for the $f_{\rm H_2}$ discrepancy between the observations and theoretical studies from an observational point of view.
The observational studies essentially tend to be biased towards bright objects such as gas-rich objects and/or optically bright objects if the targets are selected based on large optical surveys.
It has been claimed that the models reproduce the observational results relatively well 
if they take into account the sample selection criteria and observational sensitivities \citep{Kauffmann:2012ao,Popping:2014zm}.
These studies showed that the predicted $f_{\rm H_2}$ value approaches the observed value if these observational conditions are considered, but there still seem to be discrepancies between observations and model predictions \citep[see Fig. 11 in][]{Popping:2014zm}.
In addition, a gap between a semi-empirical model, as a proxy for observations, and theoretical predictions was also reported \citep{Popping:2015fu}.
This semi-empirical model couples a halo abundance matching model with a pressure-based H$_2$ formation model and is not affected by the observational limitations.
However, we must pay attention to this observational limitation when comparing the observations and theoretical studies, especially for the less massive and higher-$z$ galaxy samples, which are expected to be more biased to bright (gas-rich) objects compared to massive and low-$z$ galaxy samples.

In addition to the observational bias, the uncertainty in the CO-to-H$_2$ conversion factor, $\alpha_{\rm CO}$, 
also contributes an additional error in the $f_{\rm H_2}$ estimate.
\cite{Narayanan:2012rp} formulated $\alpha_{\rm CO}$ as a function of the gas-phase metallicity and 
luminosity-weighted CO intensity.
\citet{Narayanan:2012ix} showed that the observed $f_{\rm mol}$ value with Narayanan's $\alpha_{\rm CO}$ exhibits better agreement with the predicted values obtained with cosmological simulations.
Even though the dispersion is reduced, certain systematic stellar-mass-dependent residuals still remain \citep[see Fig. 3 in][]{Narayanan:2012ix}.
Therefore, the observational causes alone may not be able to explain the discrepancy.

\section{Implications for feedback models}
\label{sec:DiscrepancyPossibleCausesModels}

As described in section \ref{sec:ComparisonObservationTheory}, there is a gap between the observed $M_\star-f_{\rm H_2}$ relations and theoretically predicted relations. 
Although some fraction of this gap can be attributed to observational uncertainties such as the selection bias and variable $\alpha_{\rm CO}$, the gap still remains.
This suggests that the gap might originate in the uncertainties of the theoretical models. 
In this section, we focus on the uncertainties of the feedback processes in the theoretical models\footnote{
Recent studies on galaxy formation based on the semi-analytic approach have suggested that the $f_{\rm H_2}$ gap between the observations and theoretical predictions decreases if the timescales for re-accretion of the ejected gas have a mass- and time-dependence \citep{Henriques:2013ff,Henriques:2015ec,Mitchell:2014fe,White:2015lh}.
Even though the physical origin for this treatment is not clear, some feedback processes may be attributed to changing the re-accretion timescale.
}.
The free parameters for the feedback processes are fixed to reproduce the observed SMF even though the SMF is not a direct outcome of the feedbacks alone, but a consequence of the several processes involved in galaxy evolution.
This is because there is no definitive observational constraint for each feedback process.
There might also be room for improvement in the H$_2$ and star formation models in terms of the molecule fraction and the K-S law, but they are set to meet the direct observational constraints for these processes.

To investigate the stellar mass and redshift dependence of each feedback process considered in theoretical studies, we construct an equilibrium model (EM) based on the study by \citet{Dave:2012zp}.
In the following subsections, we briefly describe the EM (section \ref{EM}), and subsequently, we examine the implications of the feedback processes through comparisons between the EMs and observations (section \ref{sec:ComparisonEM}).

\subsection{Equilibrium model for galaxy formation}
\label{EM}

The EM model is based on the idea that the simulated galaxies evolve in a quasi-equilibrium state \citep{Bouche:2010kq,Dave:2012zp}:
\begin{equation}
 \dot{M}_{\rm in} = \dot{M}_\star + \dot{M}_{\rm out},
 \label{eq:equilibrium}
\end{equation}
where $\dot{M}_{\rm in}$, $\dot{M}_{\rm out}$, and $\dot{M}_\star$ denote the mass inflow rate, mass outflow rate, and SFR, respectively.
This model assumes that the stochastic variations in $\dot{M_{\rm in}}$, arising due to factors such as mergers, 
are prone to return galaxies to equilibrium \citep[see][for details]{Dave:2012zp}.

\subsubsection{Stellar mass and gas fraction}

In this section, we briefly introduce the formulation of stellar mass and gas fraction in the EM according to \citet{Dave:2012zp}.
Integrating the equation $\dot{M}_\star =  \dot{M}_{\rm in} - \dot{M}_{\rm out}$,
we obtain the stellar mass $M_\star(M_{\rm h},z)$ as a function of the redshift $z$ and halo mass $M_{\rm h}$.
The outflow rate is given as
\begin{equation}
\dot{M}_{\rm out} = \eta \dot{M}_\star,
\label{eq:outflow}
\end{equation}
where the mass loading factor $\eta$ is described as $\eta = (M_{\rm h}/10^{12}~{\rm M_\odot})^{-1/3}$ \citep{Dave:2011rm,Dave:2011zl}.
The inflow rate into a galaxy $\dot{M}_{\rm in}$ is expressed as 
\begin{equation}
\dot{M}_{\rm in} = \dot{M}_{\rm grav} - \dot{M}_{\rm prev} + \dot{M}_{\rm recyc}, 
\label{eq:inflow}
\end{equation}
where $\dot{M}_{\rm grav}$, $\dot{M}_{\rm prev}$, and $\dot{M}_{\rm recyc}$ denote the baryonic inflow rate into a galaxy's halo, rate at which material ends up in the gaseous halo of the galaxy, and return infall rate, respectively.
These are expressed as
\begin{eqnarray}
&&\dot{M}_{\rm grav} = f_{\rm b} \dot{M}_{\rm h}, \\
&&\dot{M}_{\rm prev} = (1-\zeta) \dot{M}_{\rm grav}, \\
&&\dot{M}_{\rm recyc} = \frac{\alpha_{\rm Z}}{1-\alpha_{\rm Z}} \zeta \dot{M}_{\rm grav}.
\end{eqnarray}
Here, $\dot{M}_{\rm h}$ denotes the halo mass growth rate as given by
\begin{equation}
\dot{M}_{\rm h} = 25.3 \left(\frac{M_{\rm h}}{10^{12}~{\rm M_\odot}}\right)^{1.1}
(1+1.65z)\sqrt{\Omega_{\rm m}(1+z)^3+\Omega_{\Lambda}}, 
\end{equation}
which is derived from cosmological $N$-body simulations \citep{Fakhouri:2010sf},
$f_{\rm b}$ represents the cosmic baryon fraction (0.17), 
$\Omega_{\rm m}$ the matter density parameter (0.3), $\Omega_\Lambda$ the dark energy density parameter (0.7),
$\zeta$ the preventive feedback parameter (a product of four feedback parameters, $\zeta_{\rm PHOTO}\zeta_{\rm GRAV}\zeta_{\rm QUENCH}\zeta_{\rm WIND}$; see the next subsection for explanations on each feedback), 
and $\alpha_{\rm Z}$ the metallicity ratio of inflow gas and ISM\footnote{\citet{Dave:2012zp} used $\alpha_{\rm Z} = (0.5-0.1z)(M_\star/10^{10}~{\rm M_\odot})^{0.25}$ but we use $\alpha_{\rm Z} = (0.3-0.1z)(M_\star/10^{10}~{\rm M_\odot})^{0.25}$ to prevent the results from being diverged. This difference is because some of the equations, which are not fully based on physics or ab initio simulations, were not presented in \citet{Dave:2012zp} (such as $M_{\rm q}(z)$ for equation (\ref{eq:preventiveFB_quench}) and $M_{\rm wind}(z)$ for equation (\ref{eq:preventiveFB_wind})); hence, we must formulate them by ourselves.} $\alpha_{\rm Z} = (0.3-0.1z)(M_\star/10^{10}~{\rm M_\odot})^{0.25}$ \citep{Dave:2012zp}.
Substituting these equations into equation (\ref{eq:inflow}) yields 
\begin{equation}
\dot{M}_{\rm in} = \frac{\zeta}{1-\alpha_{\rm Z}} \dot{M}_{\rm grav}.
\label{eq:inflow2}
\end{equation}
If we substitute equations (\ref{eq:outflow}) and (\ref{eq:inflow2}) into equation (\ref{eq:equilibrium}), 
the SFR can be written as 
\begin{equation}
\dot{M}_\star = \frac{\zeta \dot{M}_{\rm grav}}{(1+\eta)(1-\alpha_{\rm Z})}.
\end{equation}

In the EM, if $M_\star$ and $\dot{M}_\star$ are known via the procedure described above, 
the gas fraction of a galaxy is given by
\begin{equation}
 f_{\rm gas} \equiv \frac{M_{\rm gas}}{M_{\rm gas} + M_\star} 
 = \frac{1}{1+(t_{\rm dep}\dot{M}_\star/M_\star)^{-1}},
\end{equation}
where $t_{\rm dep} \equiv M_{\rm gas}/\dot{M}_\star$ denotes the depletion time of the gas component by star formation. 
We use the Hubble time $t_{\rm H}$ to parameterise $t_{\rm dep} = 0.4 t_{\rm H} (M_\star/10^{10}~{\rm M_\odot})^{-0.3}$ 
as discussed in \citet{Dave:2012zp}.

\subsubsection{Preventive feedback parameters}

\begin{figure*}
\includegraphics[width=150mm]{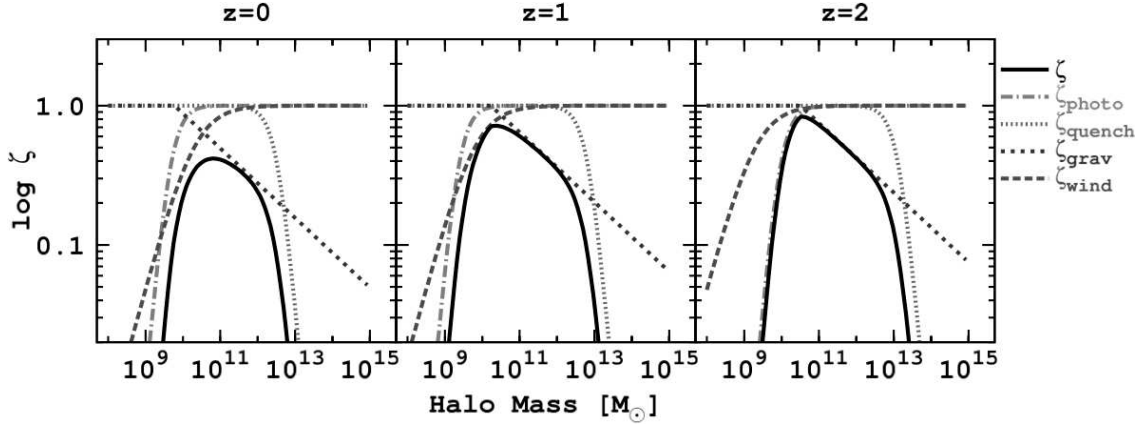}
\caption{
Preventive feedback parameters as a function of halo mass at $z =$ 0, 1.0, and 2.0.
The total preventive feedback parameter $\zeta$ (black solid line) is the product of 
four preventive feedback parameters, $\zeta_{\rm PHOTO}$ (gray dashed lines), $\zeta_{\rm GRAV}$ (gray dotted lines),
$\zeta_{\rm QUENCH}$ (gray short-dotted lines), and $\zeta_{\rm WIND}$ (gray dashed lines).
}
\label{fig:EMmodel}
\end{figure*}

Each preventive feedback parameter\footnote{In Dave's model, all feedbacks are modeled in `preventive' manner but some are represented as `ejective' feedback.} $\zeta$ depends on the halo mass $M_{\rm h}$ and redshift $z$.
The EM considers four feedback processes as follows:
\begin{itemize}
\item $\zeta_{\rm PHOTO}$, photoionising heating (hereafter PHOTO): 
An ultraviolet background radiation heats the halo gas of galaxies.
It effectively works in less massive systems since less massive galaxies cannot confine or 
acquire heated (photoionised) gas due to their shallow potential wells \citep[e.g.][]{Gnedin:2000vn,Okamoto:2008rz}.
\begin{equation}
 \zeta_{\rm PHOTO} = \left[1+ \frac{1}{3}\left(\frac{M_\gamma(z)}{M_{\rm halo}}\right)^2\right]^{-1.5},
\end{equation}
where the photosuppression mass, $M_\gamma(z)$ is taken from Fig. B1 of \citet{Okamoto:2008rz}.

\item $\zeta_{\rm GRAV}$, virial shock heating during gas accretion (hereafter GRAV): 
It effectively works in massive systems since the sparsity and high virial temperature of the halo gas in massive systems prevent the halo gas from accreting onto the galaxy \citep[e.g.][]{Faucher-Giguere:2011kk}.
\begin{equation}
 \zeta_{\rm GRAV} = {\rm MIN}\left[1,0.47\left(\frac{1+z}{4}\right)^{0.38} \left(\frac{M_{\rm halo}}{10^{12}~\rm M_\odot}\right)^{-0.25} \right].
\end{equation}

\item $\zeta_{\rm QUENCH}$, heating up the ISM and/or halo gas and preventing cooling flow from the halo due to supermassive black hole (BH) growth (or AGN feedback; hereafter QUENCH): 
This feedback works in massive systems \citep{Croton:2006rt,Gabor:2011fk}.
\begin{equation}
 \zeta_{\rm QUENCH} = \left[1 + \frac{1}{3}\left(\frac{M_{\rm halo}}{M_{\rm q}(z)}\right)^2\right]^{-1.5},
\label{eq:preventiveFB_quench}
\end{equation}
where the quenching mass is given by $M_{\rm q}(z) = 10^{12 + 0.3(1+z)}~\rm M_\odot$.
We formulate $M_{\rm q}(z)$ so as to represent the red line in Fig. 1 of \cite{Dave:2012zp}.

\item $\zeta_{\rm WIND}$, heating of surrounding gas by SNe (hereafter WIND): 
it effectively works in less massive systems since ubiquitous SNe wind tends to blow out the gas 
in less massive galaxies due to their shallow potential wells \citep[e.g.][]{Oppenheimer:2010wq}.
\begin{equation}
 \zeta_{\rm WIND} =  \left[1 + \frac{1}{50}\left(\frac{M_{\rm wind}(z)}{M_{\rm halo}}\right)\right]^{-1},
\label{eq:preventiveFB_wind}
\end{equation}
where the wind mass is given by $M_{\rm wind} = 10^{12.5-0.5(1+z)}~\rm M_\odot$.
We formulate $M_{\rm wind}$ so as to represent the blue line in Fig. 1 of \cite{Dave:2012zp}.
\end{itemize}
We present the preventive feedback parameters as a function of halo mass in Fig. \ref{fig:EMmodel} \citep[see also Fig.1 of][]{Dave:2012zp}.

\subsection{Comparisons between the EMs and observations}
\label{sec:ComparisonEM}

\begin{figure*}
\includegraphics[width=180mm]{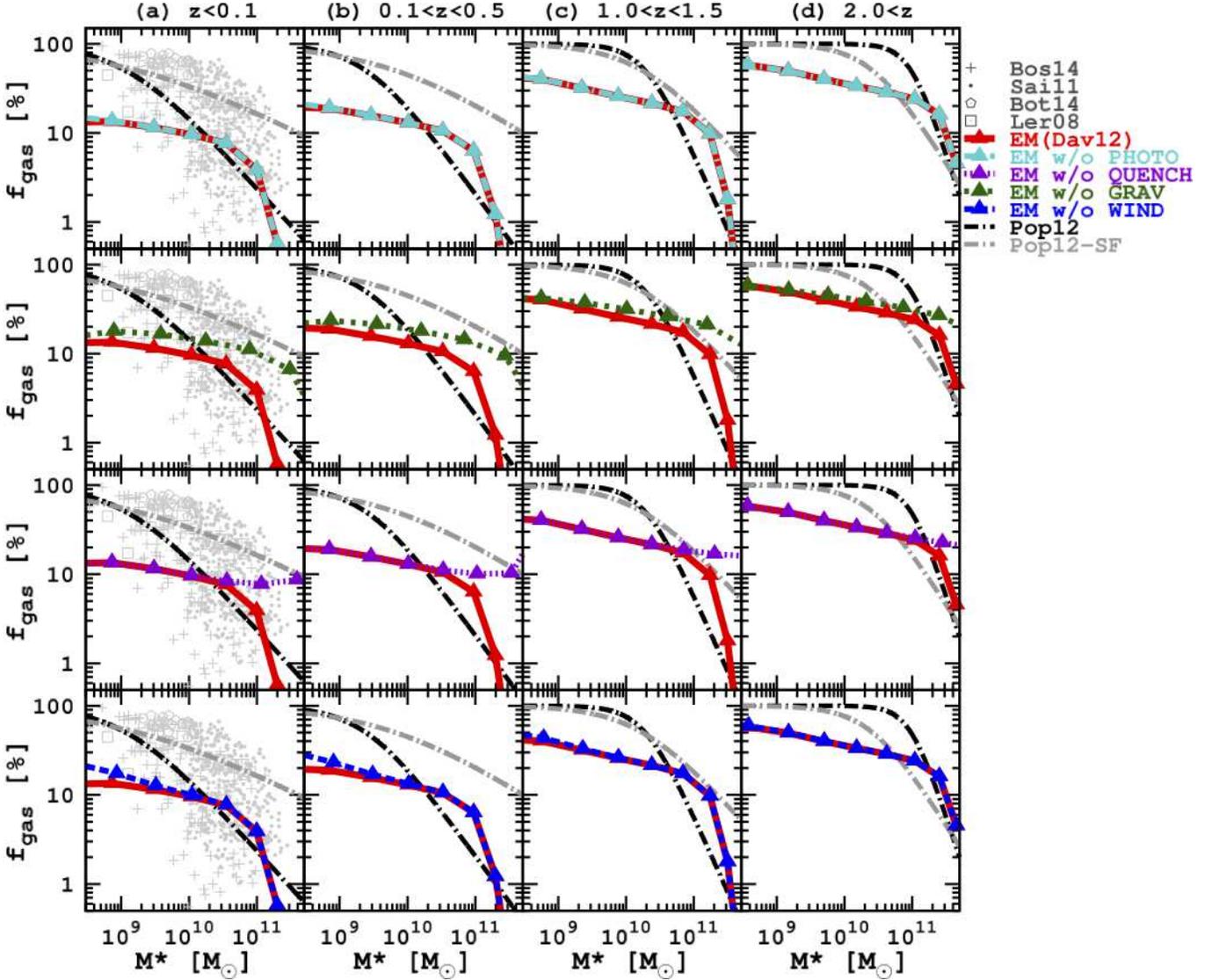}
 \caption{Total gas fraction $f_{\rm gas}$ as a function of stellar mass and redshift (a: $z<0.1$, b: $0.1<z<0.5$, c: $1.0<z<1.5$ and d: $2.0<z$).
 The black and grey dot-dash lines indicate the $M_\star-f_{\rm gas}$ relation based on the K-S law for whole and star-forming samples in \citet{Popping:2012gd}.
The redshifts of the equations for $f_{\rm mol, P12, SF}$ and $f_{\rm mol, P12}$ are 0.025 for (a), 0.3 for (b), 1.25 for (c), and 2.2 for (d).
 The lines with filled triangles indicate the results of the `equilibrium model (EM)' by \citet{Dave:2012zp} (red solid lines: model with all feedbacks, light-blue dot-dashed lines: model without photoionising feedback (PHOTO), green dotted lines: model without shock-heating during accretion (GRAV), purple short-dotted lines: model without quenching associated with black hole growth (QUENCH), and blue dashed lines: model without additional feedback associated with wind (WIND)).
 On each plot, we show the result of EM calculations at $z=0$ in (a), $z=0.2$ in (b), $z=1$ in (c) and $z=2$ in (d).
The $M_\star-f_{\rm gas}$ relations from the direct observations (HI and CO) are plotted as a reference \citep{Leroy:2008fb,Saintonge:2011hl,Boselli:2014yq,Bothwell:2014lr}.
 }
  \label{fig:mstar_fmol_AllFB3}
\end{figure*}

We compare the observed $M_\star-f_{\rm gas}$ relations with the predictions from the EMs in Fig. \ref{fig:mstar_fmol_AllFB3}.
Here, we use Popping's $M_\star-f_{\rm gas}$ relation as the observational reference (Popping et al. in prep.), which is described below:
\begin{equation}
f_{\rm gas,P12}=\frac{1}{\exp{(\log{M_\star}-E)/F}+1},
\end{equation}
where
\begin{eqnarray}
E=9.04\left(1+\frac{z}{1.76}\right)^{0.24},~~
F=0.53(1+z)^{-0.91}.
\end{eqnarray}
We use the $M_\star-f_{\rm gas}$ relation because Dave's EMs do not consider H$_2$ formation.
To clarify the contribution from each feedback process, five conditions are considered: 
1) all feedback processes are considered, 
2) PHOTO is not considered, 
3) GRAV is not considered, 
4) QUENCH is not considered, and 
5) WIND is not considered.

From Fig. \ref{fig:mstar_fmol_AllFB3}, we can confirm the stellar-mass dependence and observe the degree of influence of each feedback process as expected from Fig. \ref{fig:EMmodel}.
For less massive galaxies (but $>10^{9.2}$ M$_\odot$ according to observational completeness), the theoretically predicted $f_{\rm gas}$ value approaches the observation value at $z=0$ (Fig. \ref{fig:mstar_fmol_AllFB3}a), but it is considerably lower than the observation value even if we exclude the PHOTO and WIND feedback processes.
On the other hand, the QUENCH and GRAV feedback processes work well for massive galaxies.
The stellar-mass dependence of $f_{\rm gas}$ in the high-mass range (but $<10^{11}$ M$_\odot$ according to observational completeness) appears to be different in that there is a knee around $M_\star \sim10^{10.5}$ M$_\odot$ in the EM as opposed to a monotonic decrease in observations; however, it is not conclusive because these are consistent within the typical dispersion of the observed relation.

From Fig. \ref{fig:mstar_fmol_AllFB2}, we observe that theoretical studies tend to predict lower gas fractions than observations, particularly in less massive and massive galaxies.
Our result suggests that the feedback models implemented in the theoretical studies are clearly required.
However, the physical ingredients in these models are still insufficient to correctly reproduce the available observations.
One requirement is to consider a feedback model that quenches star formation but retains the cold-gas component in galaxies.

The EM predicts a stellar-mass dependence of $f_{\rm gas}$ similar to that of $f_{\rm H_2}$ predicted by the SAMs, rather than the SIMs; there is a kink at the characteristic stellar mass.
The $M_\star-f_{\rm gas}$ relations predicted by the EMs without $\zeta_{\rm GRAV}$ or $\zeta_{\rm QHENCH}$ become flat, similar to the predictions of the SIMs.
The observations indicate that the $M_\star-f_{\rm gas}$ or $M_\star-f_{\rm H_2}$ relations are similar to the predictions of the EM and SAMs at higher redshifts, while these relations are similar to the SIMs at lower redshifts.
The virial shock heating is expected to be included in the SIMs naturally, but AGN feedback is implemented as an ad-hoc model. 
If the observed $f_{\rm gas}$ evolution in the high-mass range of \citet{Popping:2012gd} is correct, our comparison suggests a more drastic redshift evolution of AGN feedback than the SIMs, i.e. a stronger AGN feedback for a high redshift and a weaker one for a low redshift.
However, the details regarding the modelling of AGN feedback are beyond the scope of this paper.

Some recent studies have emphasised the relevance of the feedback processes, which suppress star formation while simultaneously preserving the cold gas \citep{Makiya:2014qy,Morokuma-Matsui:2015ij}.
Our results showed that the EMs and the SAMs tend to underestimate $f_{\rm gas}$ or $f_{\rm H_2}$ in less massive and massive galaxies.
There is a larger gap between the observations and theoretical predictions for less massive galaxies in Fig.~\ref{fig:mstar_fmol_AllFB3}a ($M_\star-f_{\rm gas}$ relation) than in Fig.~\ref{fig:mstar_fmol_AllFB2}a ($M_\star-f_{\rm H_2}$ relation).
This larger gap in $f_{\rm gas}$ in less massive galaxies suggests a lot of ejection of cold gas by SN feedback modeled in the theoretical studies.
The gap may be reduced if we consider the suppression of H$_2$ formation instead of the ejection of total cold gas.

\citet{Makiya:2014qy} implemented a star formation law with a feedback depending on galaxy-scale mean dust opacity and metallicity in SAM and succeeded in reproducing the faint-end slopes of galaxy luminosity functions at $z=0$ with a {\it reasonable} strength of the SN feedback.
In their model, star formation in less massive galaxies is suppressed because such galaxies tend to have lower metallicity, lower surface density, and thus, lower H$_2$ fraction.
For massive galaxies, morphological quenching is one of the important processes, which can quench star formation even in the presence of a gas component.
\citet{Martig:2009qy} showed that star formation is suppressed once the central spheroidal component becomes sufficiently massive to stabilise the galactic disc against local gravitational instability.
Some observational studies suggested that morphological quenching is in action in the early-type galaxies \citep{Martig:2013fr,Morokuma-Matsui:2015ij}.

\section{Summary}
\label{sec:Summary}

Galaxy formation models are constructed so as to reproduce the observed SMF of galaxies 
by blowing out and/or heating up the cold-gas component to quench further star formation, particularly in less massive and massive systems.
To investigate whether the current galaxy formation models also reproduce the observed properties of the cold-gas component of galaxies, 
we compare the observed and theoretically predicted evolution of 
the molecular gas fraction with respect to the total baryonic mass as a function of stellar mass (i.e. the $M_\star-f_{\rm mol}$ or $M_\star-f_{\rm H_2}$ relations).
Our main findings are as follows:

\begin{enumerate}
\item The evolution of the observed $M_\star-f_{\rm mol}$ relation shows a stellar-mass dependent evolution;
massive galaxies have largely depleted their molecular gas at $z=1$, whereas less massive galaxies tend to convert molecular gas into stars in the regime of $0<z<1$.

\item The current cosmological hydrodynamic simulations and semi-analytical models for galaxy formation 
succeed in `qualitatively' reproducing the $M_\star-f_{\rm H_2}$ relation evolution in the regime of $0<z<2$
where less massive galaxies are always more gas-rich than massive galaxies.

\item There is a non-negligible gap between the observed and theoretically predicted $M_\star-f_{\rm H_2}$ relations.
The predicted $f_{\rm mol}$ value is more than two times smaller than the observed value at most, particularly in less massive ($<10^{10}$ M$_\odot$) and massive galaxies ($>10^{11}$ M$_\odot$).
Accordingly, for this gap, the use of feedback models could be required, which quench star formation but retain the cold-gas component.

\item The observed redshift evolution of the cold gas mass fraction in massive galaxies suggests a more drastic evolution of AGN feedback than the one implemented in the hydrodynamic simulations, i.e. a stronger AGN feedback in higher redshift ($z>1$) and a weaker one in lower redshift ($z<1$).

\item The SMF of galaxies alone is not sufficient for constraining the galaxy formation models and the $M_\star-f_{\rm mol}$ relation forms one of the important observational constraints on feedback processes implemented in galaxy-formation models.

\end{enumerate}
It is to be noted that we adopted the indirect estimates of $f_{\rm mol}$ using the optical survey data \citep{Popping:2012gd} for comparison with the galaxy formation models in this study.
To form a definitive conclusion regarding the $f_{\rm mol}$ evolution of galaxies (particularly less massive and massive galaxies), it is necessary to directly measure the cold gas mass of galaxies over wide ranges of stellar mass and redshift.
The mass function of the stellar component is of course a basic observed constraint, which must be satisfied for the models.
The cold-gas component is also an important component in the baryonic cycle of galaxies.
A combination of the observational constraints on stellar and cold gas components will aid in sketching a complete scenario of galaxy formation and evolution.
Observations with forthcoming facilities such as the ALMA, the NOEMA, and the SKA, will allow us to explore the statistical properties and evolution of cold-gas components in galaxies, such as ${\rm H_I}$ and H$_2$ mass functions, as well as the $M_\star-f_{\rm mol}$ relation.

\section*{Acknowledgments}

We thank the referee, Gerg$\ddot{\rm o}$ Popping, for a very thorough report which led to a substantial improvement of the paper.
JB was supported by JSPS Grant-in-Aid for Young Scientists (B) Grand Number 26800099.
We would like to thank Takeshi Okamoto, Masahiro Nagashima and Ryu Makiya for productive discussions, and Ross Burns, Daisuke Iono and Editage (www.editage.jp) for English language editing.

Funding for SDSS-III has been provided by the Alfred P. Sloan Foundation, the Participating Institutions, the National Science Foundation, and the U.S. Department of Energy Office of Science. The SDSS-III web site is (http://www.sdss3.org/). SDSS-III is man- aged by the Astrophysical Research Consortium for the Participating Institutions of the SDSS-III Collaboration including the University of Arizona, the Brazilian Participation Group, Brookhaven National Laboratory, Carnegie Mellon University, University of Florida, the French Participation Group, the German Participation Group, Harvard University, the Instituto de Astrofisica de Canarias, the Michigan State/Notre Dame/JINA Participation Group, Johns Hopkins University, Lawrence Berkeley National Laboratory, Max Planck Institute for Astrophysics, Max Planck Institute for Extraterrestrial Physics, New Mexico State University, New York University, Ohio State University, Pennsylvania State University, University of Portsmouth, Princeton University, the Spanish Participation Group, University of Tokyo, University of Utah, Vanderbilt University, University of Virginia, University of Washington, and Yale University.


\appendix

\bsp

\label{lastpage}

\end{document}